\documentstyle {article}

\begin{document}

{\bf Comparison of the fluctuation influence on the resistive properties of
the mixed state of $Bi_{2}Sr_{2}Ca Cu_{2}O_{8+x}$ and of thin films of
conventional superconductor.}

\

$A.V.Nikulov^{1)}, \ E.Milani^{2)}, \ G.Balestrino^{2)}, \ and \
V.A.Oboznov^{3)}$

\

$^{1)}$Institute of Microelectronics Technology and High Purity
Materials, Russian Academy of Sciences, 142432 Chernogolovka, Moscow
District, Russia.

$^{2)}$INFM, Dipartimento di Scienze e Tecnologie Fisiche ed Energetiche,
Universita di Roma "Tor Vergata", I-00133 Roma, Italy

$^{3)}$Institute of Solid State Physics, Russian Academy of Sciences,
142432 Chernogolovka, Moscow District, Russia.

\

The resistive properties of layered HTCS $Bi_{2}Sr_{2}Ca Cu_{2}O_{8+x}$ in
the mixed state are compared with those of thin films of conventional
superconductors with weak disorder (amorphous $Nb_{1-x}O_{x}$ films) and
with strong disorder ($Nb_{1-x}O_{x}$ films with small grain structure).
The excess conductivity in the mixed state is considered as a function of
the superconducting electron density and the phase coherence length.  It is
shown that the transition into the Abrikosov state differs from the ideal
case both in $Bi_{2}Sr_{2}Ca Cu_{2}O_{8+x}$ and $Nb_{1-x}O_{x}$ films, i.e.
the appearance of the long-range phase coherence is a smooth transition in
both cases. The quantitative difference between thin films with weak and
strong disorder is greater than the difference between
$Bi_{2}Sr_{2}CaCu_{2}O_{8+x}$ and conventional superconductors, showing
that the dimensionality of the system, rather than the critical
temperature, is the key factor ruling fluctuation effects.

\

I. INTRODUCTION

\

One of the obstacles to the wide application of high temperature
superconductors (HTSC) in a high magnetic field is the fluctuation
influence on the resistive properties of the mixed state. Therefore this
topic is much more popular now \cite{blatter} than it was before the
discovery of HTSC. Previously, the majority of scientists thought that the
fluctuation effects in superconductors are quite small, and now many are
of the opinion that the fluctuation effects are big in HTSC only
\cite{fisher}. Indeed, it is currently believed that higher critical
temperatures enhance drastically the effects of thermal fluctuations.
However, this is not right.

The fluctuation value is determined by the ratio of the thermal energy,
$k_{B}T$, to the effective energy of a superconductor $F_{eff} =
(f_{n}(T)-f_{s}(T))V_{ef}$. Here $f_{n}(T)-f_{s}(T)$ is the difference of
the energy density in the normal and superconducting states; $V_{ef}$ is
the effective volume. The fluctuation effects are small if
$|k_{B}T/F_{eff}| \ll 1$. Since at $T = T_{c}$, $f_{n}(T)-f_{s}(T) = 0$ a
critical region exists near the critical temperature where the fluctuation
effects are not small. The width of the critical region can be estimated by
the relation $|k_{B}T/(f_{n}(T)-f_{s}(T))V_{ef}| \simeq 1$.

According to the Ginzburg-Landau theory $f_{n}(T)-f_{s}(T) =
(1-t)|1-t|H_{c}^{2}(0)/8\pi$ \cite{huebener}, where $t = T/T_{c}$ and
$H_{c}(0)$ is the thermodynamic critical field. The effective volume in
bulk (three-dimensional) superconductors is $V_{ef} = \xi^{3}(T)$,
$\xi(T)$ being the coherence length (the effective volume of anisotropic
superconductors is $V_{ef} = \xi_{x}\xi_{y}\xi_{z}$). If one dimension of
the superconductor $d < \xi(T)$ then the effective volume is $V_{ef} =
d\xi^{2}(T)$ and the superconductor can be considered as two-dimensional.
The effective volume of a one-dimensional superconductor (two dimensions of
which are smaller than the coherence length) is $V_{ef} =
d_{x}d_{y}\xi(T)$, and for a zero-dimensional superconductor $V_{ef} =
d_{x}d_{y}d_{z}$.

According to the Ginzburg-Landau theory $\xi(T)= \xi(0)|1-t|^{-1/2}$, so
that the width $|1-t|_{cr}$ of the critical region is given by the
following Ginzburg numbers for the relevant dimensionality of the system:

$Gi_{3D} = (k_{B}T_{c}/H_{c}^{2}(0)\xi^{3}(0))^{2}$

$Gi_{2D} = k_{B}T_{c}/H_{c}^{2}(0)d \xi^{2}(0)$

$Gi_{1D} = (k_{B}T_{c}/H_{c}^{2}(0)d_{x}d_{y}\xi(0))^{2/3}$

$Gi_{0D} = (k_{B}T_{c}/H_{c}^{2}(0)d_{x}d_{y}d_{z})^{1/2}$.

The value of the Ginzburg number determines the strength of the thermal
fluctuations. $T_{c}/H_{c}^{2}(0) \simeq (dH_{c}/dT)^{-2}T_{c}^{-1}$. The
$dH_{c}/dT$ value of HTSC is not bigger than the one of conventional
superconductors \cite{salamon}. Consequently, higher critical temperatures
can not be the reason for the enhanced thermal fluctuations in HTSC. The
cause of this enhancement are the shorter coherence length and the
quasi-two-dimensionality of many HTSC.

Being $k_{B}T_{c}/H_{c}^{2}(0)\xi^{3}(0) < 1$ for all superconductors, the
Ginzburg number increases with decreasing dimensionality. As it was shown
first in \cite{lee72} the effective dimensionality of fluctuation reduces
to two near the second critical field, $H_{c2}$, in a region where the
lowest Landau level (LLL) approximation is valid. (The LLL approximation is
valid at $h \gg Gi$, $h-h_{c2} \ll 2h$ and $h > h_{c2}/3$, where $h =
H/H_{c2}(0)$, $h_{c2} = H_{c2}(T)/H_{c2}(0)$, $H_{c2}(0)=
-T_{c}(dH_{c2}/dT)_{T=T_{c}}$). Therefore, the fluctuation effects increase
in high magnetic field. The Ginzburg number in the LLL approximation is
$Gi_{3D,H} = Gi_{3D}^{1/3}(th_{c2})^{2/3}$ for three-dimensional
superconductors and $Gi_{2D,H} = Gi_{2D}^{1/2}(th_{c2})^{1/2}$ for two-
dimensional superconductors. Consequently, the difference among the
fluctuation effect values in HTSC and conventional superconductors
decreases in high magnetic field.

Thus, the fluctuation values in the mixed state of HTSC and conventional
superconductors do not differ very much and therefore the fluctuation
effects in these superconductors can be compared. Such comparison
\cite{nik96} has shown that there is no reason to think that the
fluctuation effects in $YBa_{2}Cu_{3}O_{7-x}$ qualitatively differ from
those in bulk conventional superconductors.

In the present work we compare the fluctuation effects in quasi-two-
dimensional HTSC ($Bi_{2}Sr_{2}Ca_{2}Cu_{3}O_{10+x}$ films) and in thin
films of conventional superconductor ($Nb_{1-x}O_{x}$). One ought expect
that the fluctuation effects of two-dimensional superconductors does not
depend on the $\xi(0)$ value in a enough high magnetic field, because the
fluctuation is zero-dimensional in the LLL approximation. $Gi_{2D,H} =
(Gi_{2D}th_{c2})^{1/2} = (k_{B}T/H_{c}^{2}(0)d \xi^{2}(T))^{1/2} \simeq
(k_{B}T2\pi H/H_{c}^{2}(0)\Phi_{0})^{1/2}$ at $H \simeq H_{c2}(T)$.
Where $\Phi_{0} = hc/2e$ is the flux quantum. We use the relation
$\Phi_{0} = 2\pi H_{c2}\xi^{2}$ \cite{huebener}.

The long-range order of a superconducting state is the long-range phase
coherence. Two main fundamental aspects of superconductivity, zero
electrical resistance and the Meissner effect are conditioned by it. The
Abrikosov state is the mixed state with long-range phase coherence.
According to the mean field approximation \cite{abrikos} the long-range
phase coherence appears simultaneously with a non-zero density of
superconducting electrons, $n_{s}$, at the second order phase transition
taking place at $H_{c2}$.

But according to the fluctuation theory the appearance of long-range phase
coherence is not connected directly with the $n_{s}$ value. Therefore a
mixed state without long-range phase coherence can exist according to the
fluctuation theory. The transition from the mixed state without phase
coherence to the Abrikosov state (i.e. to the mixed state with long-range
phase coherence) can have different nature in superconductors with weak and
strong disorders \cite{nik98}. This topical problem was investigated
insufficiently. Therefore it is considered first of all in the present
work.

Some of superconductor characteristics (for example such thermodynamic
properties as specific heat \cite{hake} and magnetization \cite{nik84})
depend only on the $n_{s}$ value near $H_{c2}$. Whereas the transport
properties strongly depend on the length of the phase coherence. The
thermodynamic everage value of $n_{s}$ depends very weakly on the length of
phase coherence. Therefore the resistive properties of
$Bi_{2}Sr_{2}Ca_{2}Cu_{3}O_{10+x}$ films and $Nb_{1-x}O_{x}$ films with
weak and strong disorder are studied in our paper.

\

II. THEORETICAL CONSIDERATION

\

The resistivity has different nature in the mixed states with and without
long-range phase coherence. A direct voltage can be in a state with
long-range phase coherence only if the phase difference changes in
time. This means that the resistivity in the Abrikosov state, $\rho_{f}$,
is caused by the vortex flow. The $\rho_{f}$ is called flux
flow resistivity \cite{huebener}. This denomination is not really correct
because it is obvious that magnetic flux does not flow in a superconductor.
We will use a more correct denomination - vortex flow resistivity. It is
obvious also that the Lorentz force can not be the driving force on an
Abrikosov vortex \cite{chen} because the vortex is a singularity but not a
magnetic flux.

A resistivity value in the mixed state without phase coherence
decreases in the consequence of the paraconductivity (i.e. excess
conductivity induced by superconducting fluctuation) but the cause of the
resistivity does not differ qualitatively from the one in the normal state.
A resistive transition from the paraconductivity regime to the vortex flow
regime should take place at the appearance of long-range phase coherence.
This transition is sharp if the length of phase coherence changes by jump
and smooth if it changes continuously.

There is important a definition of the phase coherence. Commonly a coherence
length (length of phase coherence) is defined by way of an correlation
function. Behaviour of the length of phase coherence near $H_{c2}$ differs
qualitatively from the one near $T_{c}$ (i.e. in zero magnetic field) in a
consequence of the reduction of the effective dimensionality on two near
$H_{c2}$. The coherence length increases up to infinity at the second order
phase transition occurred in $T_{c}$. Whereas the transversal (across
magnetic field) coherence length changes little near $H_{c2}$ (see for
example \cite{tinkham}). In the LLL region it is equal approximately
$(\Phi_{0}/B)^{1/2}$ \cite{tinkham}. Here B is the magnetic induction.
Consequently, if we define the phase coherence by way of the correlation
function we can conclude that the long-range phase coherence can not be in
the mixed state of a type II superconductor.

On other hand it is obvious that the Abrikosov state is the mixed state
with long-range phase coherence. The Abrikosov vortices appears because
according to the relation \cite{huebener}

$$\int_{l} dR\lambda_{L}^{2}j_{s} = \frac{\Phi_{0}}{2\pi}\int_{l}dR
\frac{d\phi}{dR} - \Phi \eqno{(1)}$$
magnetic flux $\Phi$ can not penetrate in a superconductor with long-range
phase coherence and without singularities. Here $\lambda_{L} =
(mc/e^{2}n_{s})^{0.5}$ is the London penetration depth; $j_{s}$ is the
superconducting current density; l is a closed path of integration; $\Phi$
is the magnetic flux contained within the closed path of integration l.
Consequently, the definition of phase coherence by way of the correlation
function is unsuited for the mixed state. It should follow from a right
definition that the existence of Abrikosov vortices (as singularities in
the mixed state with long-range phase coherence) is evidence of phase
coherence.

Such right definition is proposed in the work \cite{nik97}. It is proposed
in \cite{nik97} to use the relation (1) for the definition of phase
coherence. The phase coherence exists on a length D if the relation (1) is
valid for closed path of integration $l \simeq \pi D$ with diameter equal
D. A maximum value of D may be considered as the length of phase coherence.
If the length of phase coherence is not greater than $(\Phi_{0}/B)^{1/2}$
then a magnetic flux penetrate within superconductor without singularities
(i.e.  without Abrikosov vortices). The absence of a singularity means that
$\int_{l}dR \frac{d\phi}{dR} = 0$ for all l. We call this state without
vortices as mixed state without phase coherence.

If the relation (1) is valid on a length considerablely greater than
$(\Phi_{0}/B)^{1/2}$ then singularities appear inside a superconducting
region. $\int_{l}dR \ d\phi/dR = 2\pi N$ if l is a closed path of
integration around N singularities. A minimum value of the energy is
corresponded to a minimum possible value of the superconducting current
density $j_{s}$. Therefore according to (1) $N\Phi_{0} \simeq \Phi$, or
$n\Phi_{0} \simeq B$, here n is the density of the Abrikosov vortices.

Because the transversal coherence length changes little near $H_{c2}$, only
two characteristic lengths ($(\Phi_{0}/B)^{1/2}$ and the sample size L)
exist in an ideal (without pinning disorder) superconductor. Therefore, the
length of phase coherence can change only by jump from the small value
$(\Phi_{0}/B)^{1/2}$ to the big value L and the transition to the
Abrikosov state should be first order in superconductors without pinning
disorders. Sharp changes of resistive properties should be observed at this
transition.

Such sharp change is indeed observed in bulk superconductors with weak
disorder at $H_{c4} < H_{c2}$ \cite{nik81l,welp}. ($H_{c4}$ marks the
position of the transition to the Abrikosov state \cite{nik90}.) The
correct interpretation of this transition was proposed first in paper
\cite{nik81l}. But in HTSC \cite{welp} this sharp change was interpreted as
the vortex lattice melting. This popular interpretation can not be right
\cite{nik99} because no transition from the vortex liquid to the mixed
state without phase coherence is observed above $H_{c4}$. Results of
\cite{shilling} allow to suppose that the transition to the Abrikosov state
of bulk superconductors with weak disorders is indeed the first order phase
transition.

Pinning disorders can change the type of the transition to the Abrikosov
state, because a distance between the pinning centers is an additional
characteristic length. The length of phase coherence changes not
by jump but continuously if a density of pinning centers is big. Therefore
the resistive properties change smoothly at the transition to the Abrikosov
state of superconductors with strong pinning disorders.

Qualitative explanation of continuous increase of the length of phase
coherence (and smooth resistive transition to the Abrikosov state) in
superconductors with strong pinning disorders is proposed in work
\cite{nik98}. It is proposed to return to the Mendelssohn's model
\cite{mendelss} and to consider a real superconductor with pinning
disorders as an intermediate case between the Mendelssohn's and Abrikosov's
models. The Mendelssohn's model is a limit case of strong pinning. The
Abrikosov's model is a limit case of weak pinning. The transition into the
Abrikosov state has different nature in the Abrikosov's model and the
Mendelssohn's model.

The length of phase coherence can change only by jump from
$(\Phi_{0}/B)^{1/2}$ to a sample size L in the Abrikosov's model, because
the effective dimensionality of the fluctuations is reduced on two (from
two to zero in a two-dimensional superconductor) in a high magnetic field.
The Mendelssohn's sponge \cite{mendelss} is a system of one-dimensional
superconductors. Magnetic field does not change the fluctuation
dimensionality in a one-dimensional superconductor. Therefore we may say
that pinning disorders increase the effective dimensionality of a
two-dimensional superconductor near $H_{c2}$ from zero to one. The
appearance of phase coherence is smooth transition in one-dimensional
superconductor \cite{grunberg}.

Thus, the transition into the Abrikosov state of a real superconductor can
be both sharp (if this superconductor has weak pinning and therefore is
close to the Abrikosov's model) and smooth (if it has strong pinning and
therefore is close to the Mendelssohn's model). It is easy to find
experimentally the position of the transition to the Abrikosov state,
$H_{c4}$, in a superconductors with weak pinning, because the sharp change
of the resistive properties is observed in this case \cite{nik81l,welp}.
But it is not so easy to determine a $H_{c4}$ position in a superconductor
with strong disorder. Because the transition to the Abrikosov state is
continuous in the Mendelssohn's model we ought consider the $H_{c4}$ in a
superconductor with strong pinning as a magnetic field (or temperature)
region (the $H_{c4}$ region) in which the length of phase coherence changes
from $(\Phi_{0}/B)^{1/2}$ to a sample size L.

A continuous transition from the paraconductivity regime to the vortex flow
regime takes place in this region. The vortex pinning exists in the vortex
flow regime both of a real type II superconductor and a Mendelssohn's
sponge \cite{nik98}.  As a consequence of pinning the vortices can not flow
when the transport current is lower than a critical current $j_{cs}$,
sometimes called static critical current, and practically measured through
a small but finite voltage level. At $j \gg j_{cs}$ the current voltage
characteristics can be described in any case \cite{kim} by the equation

$$E = V/l = \rho_{f}(j - j_{cd}) \eqno{(2)}$$
where $j_{cd}$ is called dynamic critical current.

The resistivity at a small ($j < j_{cd}$) current can have a finite value
in a consequence of the thermally activated vortex creep. Therefore $j_{cs}$
can be zero when $j_{cd} > 0$. According to the Kim-Anderson
\cite{anders64} vortex creep theory (see \cite{brandt95})

$$E=E_{0} \sinh(j/j_{0}) \eqno{(3)}$$
where $j_{0} = k_{B}T/BV_{j}l_{j}$; $V_{j}$ is jumping volume and $l_{j}$
is the jump width.

Majority of real non-Ohmic current-voltage characteristic in the
vortex flow regime can not be described completely by the relation (2) and
(3). But we can state that non-Ohmic current-voltage characteristic is
evidence of the vortex flow regime, because they are Ohmic in the
paraconductivity regime. (We do not considered here the overheating
influence and nonlinear effects, which can be observed at a high measuring
current.) We will consider the appearance of the non-Ohmic current-voltage
characteristic as the end of the transition to the vortex flow regime. The
onset of this transition is the appearance of vortices, which takes place
when the length of phase coherence surpasses $(\Phi_{0}/B)^{1/2}$.

The onset of the transition to the Abrikosov state can be detect
experimentally by means of a comparison of experimental dependence of
excess conductivity and paraconductivity dependence of the
mixed state without phase coherence. The paraconductivity of
two-dimensional superconductors in the linear (Gaussian) approximation
region above $H_{c2}$ can be expressed by the relation

$$\sigma_{fl,2D} = \frac{\sigma_{0}}{d} F_{2D}(t,h) \eqno{(4)}$$

where d is the film thickness; $\sigma_{0} = e^{2}/\hbar = 0.00024
\Omega^{-1}$. The exact $F_{2D}(t,h)$ dependencies can be calculated from
the results of the Ami-Maki work \cite{ami78}. Near $H_{c2}$, at $h-h_{c2}
\ll h_{c2}$, $F_{2D}(t,h) \simeq F(t,h-h_{c2})$. For example the
Aslamasov-Larkin contribution near $H_{c2}$ is equal

$$\sigma_{AL,2D} \simeq \frac{\sigma_{0}}{4d} \frac{t}{h-h_{c2}}
\eqno{(5)}$$
The linear approximation is valid for $h-h_{c2} \gg Gi_{H}$. Near
$H_{c2}$, in the critical region, the fluctuation interaction must be taken
into account. The Aslamasov-Larkin contribution prevails in high magnetic
field in the critical region. In the mixed state without phase
coherence $\sigma_{AL,2D}$ is determined only by the thermodynamic average
density of superconducting electrons

$$<n_{s}> = \frac{\sum n_{s}\exp(-F_{GL}/k_{B}T)} {\sum
\exp(-F_{GL}/k_{B}T)} \eqno{(6)}$$
because the length of phase coherence is approximately constant. Here

$$\frac{F_{GL}}{k_{B}T} = \sum_{q}\epsilon_{n}|\Psi_{q}|^{2} + \frac{1}{2S}
\sum_{q_{i}}V_{q_{1},q_{2}, q_{3},q_{4}}\Psi^{*}_{q_{1}}\Psi^{*}_{q_{2}}
\Psi_{q_{3}}\Psi_{q_{4}} \eqno{(7)}$$
is the relation of the Ginzburg-Landau free energy of two-dimensional
superconductor, $F_{GL}$, to the thermal energy, $k_{B}T$. We use the
expansion $\Psi(r) = V^{-1/2}\sum_{q}\Psi_{q} \psi_{q}(r)$ and a
dimensionless unit system. $\epsilon_{n} = (t+h-1+2nh)/Gi_{2D,H}$;
$V_{q_{1},q_{2}, q_{3},q_{4}} = V^{-1}\int_{V} dV
\psi^{*}_{q_{1}}\psi^{*}_{q_{2}}\psi_{q_{3}}\psi_{q_{4}}$; $q =
(n,l)$; n is a number of a Landau level; l is the index of degenerate
eigenfunctions. We used the formula $mc\alpha_{0}/\hbar e =
H_{c2}$. $\alpha = \alpha_{0}(t - 1)$ is the coefficient of the
Ginzburg-Landau theory

Only the lowest (n=0) Landau level is taken into account in the LLL
approximation. In this approximation

$$\frac{F_{GL}}{k_{B}T} = \epsilon \overline{n_{s}} + 0.5\beta_{a}
\overline{n_{s}}^{2} \eqno{(8)}$$
where $\epsilon = \epsilon_{0}$ is the distance from $H_{c2}$ ($T_{c2}$)
($t+h-1 = t-t_{c2} = h-h_{c2}$); $\overline{n_{s}} = (\int_{V} dV n_{s})/V
= \sum_{l}|\Psi_{l}|^{2} $ is the spatial average density of the
superconducting pairs, $\beta_{a} = \overline{n_{s}^{2}}/
\overline{n_{s}}^{2} = S^{-1} \sum_{q_{i}}V_{q_{1},q_{2},
q_{3},q_{4}}\Psi^{*}_{q_{1}}\Psi^{*}_{q_{2}}
\Psi_{q_{3}}\Psi_{q_{4}}/(\sum_{l}|\Psi_{l}|^{2})^{2}$ is the generalized
Abrikosov parameter.

The expression (8) is similar to the one for zero-dimensional
superconductors. A main difference consists in a difference of the
Abrikosov parameter value. The thermodynamic average of the Abrikosov
parameter $<\beta_{a}>$ changes from 2 at $\epsilon \gg 1$ to the minimum
possible value $\beta_{A} \simeq 1.16$ \cite{1964} at $ -\epsilon \gg 1$ in
two-dimensional superconductor \cite{macdonal,nik95b} and from 2 to 1 in
zero-dimensional superconductors. This difference is small. Therefore a D-2
model can be used for description of some properties in the LLL
approximation region: theoretical results obtained for zero-dimensional
superconductors can be used to describe the experimental dependencies of
two-dimensional superconductors and results for one-dimensional
superconductors can be used for the description of bulk superconductors
\cite{thouless}.

It is obvious that the thermodynamic average of $\overline{n_{s}}$ and
$\beta_{a}$ depend only on the $\epsilon$ value in this (D-2 model)
approximation, and therefore the density of the superconducting pairs is a
universal function of $(t+h-1)/Gi_{2D,H} = (t+h-1)/(thGi_{2D})^{1/2}$. This
is a scaling law of the LLL approximation.

According to the D-2 model the paraconductivity dependence of a
two-dimensional superconductor can be described by the relation

$$\sigma_{fl,2D} = \frac{\sigma_{0}t}{dGi_{2D,H}} F(\epsilon) =
\frac{\sigma_{0}t}{dGi_{2D,H}} F((h+t-1)/Gi_{2D,H}) \eqno{(9)}$$
where $F(\epsilon)$ is a universal function. This relation is a consequence
of the LLL scaling law. It is valid if the length of phase coherence does
not exceed  $(\Phi_{0}/B)^{1/2}$. Therefore in order to detect the
onset of the transition to the Abrikosov state in the LLL region we can use
the scaling law (9). The experimental dependencies of excess
conductivity deviate from the universal dependence (9) when the length of
phase coherence begins to increase (i.e.  becomes larger than
$(\Phi_{0}/B)^{1/2}$). This method is enough substantial because the
scaling law is a general consequence of the fluctuation Ginzburg-Landau
theory in the LLL approximation \cite{nik96}. The scaling of the
paraconductivity dependencies is observed both in bulk superconductors
\cite{nik83,nik85} and in thin films \cite{nik95l,kes97} with weak
disorders not only above $H_{c2}$ but also appreciably lower than
$H_{c2}$.

Thus, we can detect the onset of the transition to the Abrikosov state in
the LLL region by means of the transgression of the scaling law (9) and the
end of this transition by means of the non-Ohmic current-voltage
characteristic. This method can be especially useful for the investigation
of the phase coherence appearance in superconductors with strong disorders,
where the transition into the Abrikosov state is smooth \cite{nik98}.

\

III. SAMPLE PREPARATION AND CHARACTERISTICS

\

The $Bi_{2}Sr_{2}CaCu_{2}O_{8+x}$ films used in this paper were grown by
liquid phase epitaxy (LPE) on $NdGaO_{3}$ substrates \cite{bmmpp91}, have a
thickness of 0.5 $\mu m$, a room temperature resistivity $\rho_{ab}(300 K)
= 250 \ \mu \Omega \ cm = 25 \ 10^{-7} \Omega m$, and an extrapolated
resistance at 0 K of less than $50 \ \mu \Omega \ cm = 5 \ 10^{-7} \Omega
m$. The films are epitaxial, with the c-axis perpendicular to the substrate
and a mosaic spread slightly larger than $0.1^{o}$. Magnetic fields up to
about 1 T were applied perpendicular to the film surface by a
conventional electromagnet.

The $Nb_{1-x}O_{x}$ films were produced by magnetron sputtering of
Nb in an atmosphere of argon and oxygen. Changing the oxygen we produced
films with different oxygen contents. The transmission electron microscopy
investigation shown that the films with small oxygen contents have small
grain structure whereas the films with greater oxygen contents are
amorphous. The temperature of the superconducting transition, $T_{c}$, of
the films decreases with increasing oxygen content. For an oxygen content
$x > 0.2$ the critical temperature $T_{c} < 2 \ K$. In the present work the
amorphous films with $x \simeq 0.2$ and the films with small ($\simeq 10 \
nm$) grain structure with $x \simeq 0.08$ were used. The oxygen content was
determined by Auger analysis with a relative error 0.3. The critical
temperature of the used amorphous films is $T_{c} = 1.8 - 3 \ K$ and that
of the used films with small grain structure is $T_{c} = 5.7 \ K$.
$(dH_{c2}/dT)_{T=Tc} = - 2.2 \ T/K$ for the amorphous films and
$(dH_{c2}/dT)_{T=Tc} = - 0.6 \ T/K $ for the crystalline films. The
temperature dependence of normal resistivity of both films is weak, the
normal resistivity $\rho_{n} = 4 \ 10^{-7} \Omega m$ of the films with
small grain structure weakly decreasing as temperature decreases, and the
resistivity $\rho_{n} = 20 \ 10^{-7} \Omega m$ of the amorphous films
increasing with decreasing temperature. This change can be connected with
weak localization.

A perpendicular magnetic field up to 5 T produced by a superconducting
solenoid was used for the measurements. It was measured with a relative
error 0.0005. The temperature was measured with a relative error 0.001. The
resistivity was measured with a relative error 0.0001.

\

IV. EXPERIMENTAL RESULTS AND DISCUSSION

\

The in-plane resistive dependencies of $Bi_{2}Sr_{2}CaCu_{2}O_{8+x}$ film,
amorphous and small grain structure $Nb_{1-x}O_{x}$ films in
perpendicular magnetic fields are compared in Fig.1. In all cases no sharp
feature is observed. The resistivity value decreases smoothly with
decreasing temperature. The magnetic field has a different influence on the
resistive transition of $Bi_{2}Sr_{2}CaCu_{2}O_{8+x}$ and $Nb_{1-x}O_{x}$
films. The resistive transition of $Bi_{2}Sr_{2}CaCu_{2}O_{8+x}$ widens
only whereas the resistive transitions of $Nb_{1-x}O_{x}$ films are
displaced to lower temperature at the increase of the magnetic field value
from 0 to 1 T (Fig.1).

The transition from Ohmic to non-Ohmic behavior of the current-voltage
characteristics is also smooth. The observed change in the current-voltage
characteristics is qualitatively similar in all films (Fig.2). From the
current-voltage characteristics three different H-T regimes can be
distinguished \cite{kes93}. 1) A high-field regime with zero critical
current and Ohmic current-voltage characteristics ($j_{cs} = 0$ and $j_{cd}
= 0$). 2) An intermediate field regime in which the critical current is
zero, but the current-voltage characteristics are non-Ohmic ($j_{cs} = 0$
and $j_{cd} > 0$). 3) A low-field regime with a nondetectable voltage below
a finite static critical current ($j_{cs} > 0$ and $j_{cd} > 0$). These
three regimes are observed both in the $Bi_{2}Sr_{2}CaCu_{2}O_{8+x}$ films
and in the $Nb_{1-x}O_{x}$ films (see Fig.2). This change in the
current-voltage characteristics observed in the films and in layered HTCS
differs from the one observed in bulk superconductors with weak disorder,
where the current-voltage characteristics change by jump from Ohmic to zero
resistivity at a current value smaller than the critical current
\cite{nik81l}.

The $H/H_{c2}$ values corresponding to the intermediate field regime are
different in the $Bi_{2}Sr_{2}CaCu_{2}O_{8+x}$ films, in amorphous and small
grain structure $Nb_{1-x}O_{x}$ films. This value is highest in
$Nb_{1-x}O_{x}$ films with small grain structure and lowest in amorphous
$Nb_{1-x}O_{x}$ films (Fig.2). For example, at a reduced temperature value
$t = T/T_{c} \simeq 0.75$, the $H/H_{c2}$ values corresponding to lower and
higher boundaries of the intermediate regime are respectively $<0.0009$ and
0.008 for the amorphous $Nb_{1-x}O_{x}$ film, 0.001 and 0.02 for
$Bi_{2}Sr_{2}CaCu_{2}O_{8+x}$, 0.35 and 0.75 for $Nb_{1-x}O_{x}$ film with
small grain structure. The higher boundary of the intermediate regime
corresponds to the $H/H_{c2}$ value at which the current-voltage
characteristics become non-Ohmic: the resistivity value $\rho =
dE/dj$ at j = 0, $\rho_{j=0}$, is lower than $0.95 \rho_{j>0}$ at a high
current value. The lower boundary of the intermediate regime corresponds to
the $H/H_{c2}$ value at which $\rho_{j=0} < 10^{-11} \Omega m$.

In the intermediate field regime the $\rho_{j=0}$ value decreases more
rapidly with the decrease of temperature or magnetic field than the
$\rho_{j>0}$ value. For example, the $\rho_{j=0}$ value decreases from $3
\ 10^{-9} \Omega m$ at H = 0.01 T to $10^{-11} \Omega m$ at H = 0.001 T in
the amorphous $Nb_{1-x}O_{x}$ film at T = 1.75 K; from $4 \ 10^{-8} \Omega
m$ at H = 0.4 T to $10^{-11} \Omega m$ at H = 0.02 T in the
$Bi_{2}Sr_{2}CaCu_{2}O_{8+x}$ film at T = 60 K and from $1.4 \ 10^{-8}
\Omega m$ at H = 7 kOe to $10^{-11} \Omega m$ at H = 3 T in the
$Nb_{1-x}O_{x}$ film with small grain structure at T = 4.2 K (Fig.2).
Whereas the $\rho_{j>0}$ value changes in these cases from $3 \ 10^{-9}
\Omega m$ to $3 \ 10^{-10} \Omega m$ in the amorphous $Nb_{1-x}O_{x}$ film
(at $j = 2 \ 10^{8} \ A/m^{2}$), from $4 \ 10^{-8} \Omega m$ to $6 \
10^{-9} \Omega m$ in $Bi_{2}Sr_{2}CaCu_{2}O_{8+x}$ film (at $j = 10^{8} \
A/m^{2}$) and from $1.4 \ 10^{-8} \Omega m$ to $1.3 \ 10^{-9} \Omega m$ in
the $Nb_{1-x}O_{x}$ film with small grain structure (at $j = 6 \ 10^{8} \
A/m^{2}$). The current-voltage characteristics of the amorphous
$Nb_{1-x}O_{x}$ film at high current values can be described by the
relation (2). Whereas the current-voltage characteristics of the
$Bi_{2}Sr_{2}CaCu_{2}O_{8+x}$ film and of the $Nb_{1-x}O_{x}$ film with
small grain structure are non-linear at high j.

The low-field regime of the $Bi_{2}Sr_{2}CaCu_{2}O_{8+x}$ and the amorphous
$Nb_{1-x}O_{x}$ films corresponds to very low reduced magnetic field
$H/H_{c2}$, whereas in the $Nb_{1-x}O_{x}$ films with small grain structure
$j_{cs} > 0$ already at a high $H/H_{c2}$ value (Fig.3). The static
critical current has most value in the $Nb_{1-x}O_{x}$ film with small
grain structure Fig.3. The $j_{cs}$ values of $Bi_{2}Sr_{2}CaCu_{2}O_{8+x}$
and the amorphous $Nb_{1-x}O_{x}$ film are close but their
dependencies on the magnetic field differ qualitatively Fig.3. The critical
current of the amorphous $Nb_{1-x}O_{x}$ film decreases sharply in low
magnetic field, whereas the one of $Bi_{2}Sr_{2}CaCu_{2}O_{8+x}$ changes
weakly. For example, the critical current of the amorphous
$Nb_{1-x}O_{x}$ film at T = 1.75 K ($T/T_{c} = 1.77 K$) exceeds $2 \ 10^{9}
A/m^{2}$ at H = 0 and is lower than $10^{4} A/m^{2}$ at H = 0.002 T,
whereas the critical current of $Bi_{2}Sr_{2}CaCu_{2}O_{8+x}$ at T = 60 K
($T/T_{c} = 0.75 K$) changes from $10^{7} A/m^{2}$ at H = 0 to $0.4 \
10^{7} A/m^{2}$ at H = 0.02 T (see Fig.3 and also Fig.2).

We interpret the higher boundary of the intermediate regime as the end of
the transition to the Abrikosov state. The length of phase coherence
exceeds a sample size in the intermediate regime, but the $\rho_{j=0}$
value is not equal zero in a consequence of the thermally assisted vortex
flow resistivity (the vortex creep) \cite{brandt95}. According to Eq. (3)
$E \simeq E_{0}j/j_{0} = \rho_{TAFF}j$ at a low current $j \ll j_{0}$.
Where $\rho_{TAFF} = E_{0}/j_{0}$ is a thermally activated linear
resistivity \cite{brandt95}.

The current-voltage characteristics of the $Nb_{1-x}O_{x}$ film with small
grain structure are described very well by Eq. (3) in a wide region of
magnetic field and temperature values as shown in Fig.4. For example at T
= 4.2 K Eq. (3) describes the current-voltage characteristics in the region
from 0.2 T to 0.6 T ($H_{c2} = 0.9 T$; at H = 0.7 T the current-voltage
characteristic is Ohmic). The $E_{0}$ value changes in this region from
0.27 V/m at H = 0.6 T to $4 \mu V/m$ at H = 0.2 T. The $j_{0}$ value
changes from $0.90 \ 10^{8} A/m^{2}$ at H = 0.6 T to $0.37 \ 10^{8}
A/m^{2}$ at H = 0.2 T.

Eq. (3) describes a crossover from the thermally activated linear
resistivity to the vortex flow resistivity $\rho_{f}$. It is obvious that
this relation can be valid at $E \ll \rho_{f}j$ only. Therefore a
deviation of the experimental data from the (3) dependence is observed at
high current values (Fig.4). If $\rho_{TAFF} \simeq \rho_{f}$ Eq.
(3) can not have a validity region. The current-voltage characteristics is
close to Ohmic in this case.

We can estimate the $V_{j}l_{j} = k_{B}T/j_{0}Bd$ value from the
comparison of the experimental dependencies in Fig.4 with the dependence
(3). Since the films are thin (the film thickness $d \simeq \xi$) $V_{j}
= dS_{j}$, where $S_{j}$ is the jumping area. The comparison shows that the
$(S_{j}l_{j})^{1/3}$ value of the $Nb_{1-x}O_{x}$ films with small grain
structure is smaller than the distance between the Abrikosov vortex in the
triangular lattice $(2\Phi_{0}/3^{1/2}B)^{1/2}$. The $(S_{j}l_{j})^{1/3}$
value increases with decreasing magnetic field as well as the
$(2\Phi_{0}/3^{1/2}B)^{1/2}$ value. This means that individual vortex
creep is observed in the $Nb_{1-x}O_{x}$ films with small grain structure.

The vortex creep in the $Nb_{1-x}O_{x}$ films with small grain structure
can be described also as the creep in the Mendelssohn sponge with variable
width of superconducting threads. The vortex creep in the Mendelssohn
sponge is described by Eq.(3) at low currents \cite{nik98}. $E_{0} =
Bl_{j}\omega_{0} \exp(\xi dwf_{GL}/k_{B}T)$ and $j_{0} \simeq
8\pi^{2}k_{B}T/\xi d\Phi_{0}$ in this model. Here $\omega_{0}$ is an
attempt frequency; $f_{GL}$ is the difference of the free-energy density in
the normal and the superconducting phase; d is the film thickness; w is the
width of superconducting threads across magnetic field.

We considered a distance between the normal vortex cores in superconductors
as the width of superconducting threads across magnetic field. Because the
radius of the normal vortex core is equal approximately $\xi(T)$ and the
distances between vortex centers is equal $a(\Phi_{0}/H)^{0.5}$ then
$w(T,H) \simeq a(\Phi_{0}/H)^{0.5} - 2\xi(T) \simeq 2\xi(T)
((H_{c2}/H)^{0.5} - 1)$. a being a number of order 1 (in the triangular
lattice $a = (2/3^{0.5})^{0.5}$). Consequently, according to our model
$\ln(E_{0}/E_{0,H_{c2}}) = A((H_{c2}/H)^{1/2}-1)$. Where $E_{0,H_{c2}} =
Bl_{j}\omega_{0}$ is a fit parameter, because we do not know the
$\omega_{0}$; the $A = 2\xi^{2} df_{GL}/k_{B}T = \xi^{2}dH_{c}^{2}(0)/4\pi
k_{B}T$ value can be evaluated from known parameters of the superconducting
film. The experimental dependencies $E_{0}(H)$ of the $Nb_{1-x}O_{x}$ films
with small grain structure can be described qualitatively by this relation:
the $\ln(E_{0})$ value is proportional to $((H_{c2}/H)^{1/2}-1)$ in a wide
region of the H values. But the experimental A value differs from the one
evaluated from the parameter values of the superconducting film. For
example at T = 4.2 K the experimental value $A \simeq 16$ whereas the
theoretical one is equal 100.

The current-voltage characteristics of the amorphous $Nb_{1-x}O_{x}$ films
can be described by Eq. (3) at the higher boundary of the intermediate
regime (Fig.4). The current-voltage characteristics are similar to the
ones of the $Nb_{1-x}O_{x}$ film with small grain structure (Fig.4),
although they are observed for magnetic field values differed more than 100
times lower. For example the current-voltage characteristics of the
amorphous $Nb_{1-x}O_{x}$ film at H = 0.004 T and T = 1.75 K ($H_{c2} =
1.14 T$) is close to the one of the $Nb_{1-x}O_{x}$ film with small grain
structure at H = 0.5 T and T = 4.2 K ($H_{c2} = 0.9 T$). The
current-voltage characteristics of $Bi_{2}Sr_{2}CaCu_{2}O_{8+x}$
considerably differ from the (3) dependence (Fig.4).

According to the classical result \cite{kim} the current-voltage
characteristics at a enough high current value should be described by the
relation (2) in the Abrikosov state. But we observed a linear part at
high current value on the current-voltage characteristics only of the
amorphous $Nb_{1-x}O_{x}$ films (Fig.2). We assume the vortex pinning of
the $Nb_{1-x}O_{x}$ films with small grain structure and the
$Bi_{2}Sr_{2}CaCu_{2}O_{8+x}$ is too strong in order to observed a "free"
vortex flow at the measuring current we can use. Therefore the $\rho_{j>0}$
only of the amorphous $Nb_{1-x}O_{x}$ films can be considered as the
$\rho_{f}$.

The vortex flow resistivity, $\rho_{f}$, has been calculated in the mean
field approximation for low ($H \ll H_{c2}$) and high ($H \simeq H_{c2}$)
magnetic fields \cite{kopnin75}. The intermediate regime of the amorphous
$Nb_{1-x}O_{x}$ films corresponds to $H \ll H_{c2}$. According to the mean
field approximation $\rho_{f} = k\rho_{n}H/H_{c2}$ in low magnetic fields
\cite{kopnin75}. The observed $\rho_{j>0}(H)$ dependencies of the amorphous
$Nb_{1-x}O_{x}$ films in the intermediate and low-field regimes can be
described by this relation. The observed $\rho_{j>0}H_{c2}/\rho_{n}H$ value
is close to the k coefficient value predicted by theory \cite{kopnin75}.

The $\rho_{j>0}(H)$ dependencies do not change appreciably at the transition
to the high-field regime. At low magnetic field the resistivity $\rho =
dE/dj$ dependencies can be described by the above relation also in the
high-field regime. But this is not evidence that the high-field regime
corresponds to the Abrikosov state. The absence any sharp change of the
resistive properties in thin film with very weak pinning disorders allows
to suppose that whole the high-field regime of the amorphous
$Nb_{1-x}O_{x}$ films corresponds to the mixed state without phase
coherence \cite{nik95l}.

The resistivity $\rho = dE/dj$ values of the $Bi_{2}Sr_{2}CaCu_{2}O_{8+x}$
and the $Nb_{1-x}O_{x}$ film with small grain structure in the high-field
regime differ from the one of the amorphous $Nb_{1-x}O_{x}$ films. For
example at $t = T/T_{c} \simeq 0.75$ and at the lower boundary of the
high-field regime the relation $\rho/\rho_{n}$ is equal 0.0015 in the
amorphous $Nb_{1-x}O_{x}$ at H = 0.01 T ($H/H_{c2} = 0.0088$); 0.064 in the
$Bi_{2}Sr_{2}CaCu_{2}O_{8+x}$ film at H = 0.4 T ($H/H_{c2} = 0.02$) and
0.036 in the $Nb_{1-x}O_{x}$ film with small grain structure at H = 0.7 T
($H/H_{c2} = 0.78$).

These $\rho$ values of the $Bi_{2}Sr_{2}CaCu_{2}O_{8+x}$ exceeds the
$\rho_{f}$ value calculated in the mean field approximation. According to
theories the $\rho_{f} H_{c2}/\rho_{n}H$ value can not exceed 1
\cite{kopnin75}, whereas the $\rho H_{c2}/\rho_{n}H$ values of the
$Bi_{2}Sr_{2}CaCu_{2}O_{8+x}$ are larger than 1 in the high-field regime.
For example at t = 0.75 and H = 0.4 T, $\rho H_{c2}/\rho_{n}H = 3$. The
$\rho$ values of the $Nb_{1-x}O_{x}$ film with small grain structure are
smaller than the $\rho_{f}$ values. For example at t = 0.75 and H = 0.7 T
($H/H_{c2} = 0.78$) $\rho /\rho_{n} = 0.036H/H_{c2}$, whereas according to
the theoretical extrapolation \cite{kopnin75} the $\rho_{f}/\rho_{n}$ value
at $H/H_{c2} = 0.78$ can not be smaller than $0.3H/H_{c2}$. The large
$\rho$ values of the $Bi_{2}Sr_{2}CaCu_{2}O_{8+x}$ may be connected with
a strong contribution of the higher Landaw levels to the order parameter.
The small $\rho$ values of the $Nb_{1-x}O_{x}$ film with small grain
structure may be connected with the large length of phase coherence below
$H_{c2}$.

In order to detect the onset of the increase of the phase coherence length
the experimental dependencies of the excess conductivity should be
comparison with the D-2 model paraconductivity dependence described by the
relation (9). According to the relation (9) the function $\Delta \sigma
Gi_{2D,H}/\sigma_{0}t = F(\epsilon) = F((h+t-1)/Gi_{2D,H})$ should be
universal in the LLL region for different magnetic field values and
different superconductors if the length of phase coherence does not exceed
$(\Phi_{0}/B)^{1/2}$). This function is plotted in Fig.5. There are used
the experimental dependencies of the excess conductivity of the
$Bi_{2}Sr_{2}CaCu_{2}O_{8+x}$ film, amorphous and small grain structure
$Nb_{1-x}O_{x}$ films in different magnetic fields. Here $\Delta \sigma =
R^{-1}(H,T) - R^{-1}_{n}$, $R(H,T) = \rho(H,T)/d$ is the resistance on a
square at given H and T, $R_{n} = \rho_{n}/d$ is the normal resistance on a
square and d is the film thickness for $Nb_{1-x}O_{x}$ (d = 20 nm in both
cases) and the spacing between $CuO_{2}$ planes for
$Bi_{2}Sr_{2}CaCu_{2}O_{8+x}$ (d = 1.5 nm in this case \cite{livan97}).

The $H_{c2}$ value of $Nb_{1-x}O_{x}$ films is determined by comparison of
the experimental and theoretical paraconductivity dependencies in the
linear approximation region. The reliability of this method has been
demonstrated by the investigation of paraconductivity in bulk
superconductors. The non coincidence of the transition into the Abrikosov
state and the $H_{c2}$ line has been discovered for the first time using
this method \cite{nik81j}. Later \cite{nik84} this non coincidence was
confirmed by determination of the $H_{c2}$ value from magnetization
measurement. The $H_{c2}$ value of the $Bi_{2}Sr_{2}CaCu_{2}O_{8+x}$ film
is determined from the results of \cite{livan97}.

The scaling (9) is observed for the amorphous $Nb_{1-x}O_{x}$ films.
We compared the $\Delta \sigma Gi_{2D,H}/\sigma_{0}t = F(\epsilon) =
F((h+t-1)/Gi_{2D,H})$ dependencies for different magnetic fields
in the region $t/t_{c2} > 0.7$. In this whole region this dependencies
are close (Fig.5). The same scaling dependence was observed in a wider
region in amorphous a-NbGe films with an intermediate level of
disorder \cite{kes97}. Therefore this $F((h+t-1)/Gi_{2D,H})$ dependence
can be considered as universal for the mixed state without phase coherence
of two-dimensional superconductors in the LLL region.

The $\Delta \sigma Gi_{2D,H}/\sigma_{0}t = F(\epsilon) =
F((h+t-1)/Gi_{2D,H})$ dependencies both of the $Nb_{1-x}O_{x}$ film with
small grain structure and of $Bi_{2}Sr_{2}CaCu_{2}O_{8+x}$ differ for
different magnetic fields (Fig.5). These dependencies deviate from the
universal dependence in different directions. Consequently, different
causes upset the scaling in the $Nb_{1-x}O_{x}$ film with small grain
structure and in $Bi_{2}Sr_{2}CaCu_{2}O_{8+x}$.

The excess conductivity dependencies of $Bi_{2}Sr_{2}CaCu_{2}O_{8+x}$
deviate from the scaling one because the LLL approximation is not valid for
HTSC at the magnetic field values used. The LLL approximation is valid at
$H \gg H_{LLL} = GiH_{c2}(0)$. For conventional superconductors $H_{LLL}
\simeq 10^{-4} T$ is a very small value, whereas for
$Bi_{2}Sr_{2}CaCu_{2}O_{8+x}$ $H_{LLL} \simeq 10 T$ \cite{nik96}. Therefore
for the magnetic field values used in our work H = 0.01 - 1 T the LLL
approximation is valid for the $Nb_{1-x}O_{x}$ films whereas for
$Bi_{2}Sr_{2}CaCu_{2}O_{8+x}$ the higher Landau levels must be taken into
account.

The higher Landau levels were taken into account in the calculation
of the paraconductivity of $Bi_{2}Sr_{2}CaCu_{2}O_{8+x}$ in the Hartree
approximation in \cite{livan97}. The obtained theoretical dependencies
describe well enough the experimental dependencies in the high temperature
region of the resistive transition \cite{livan97}. This means that the
phase coherence length does not exceed considerably $(\Phi_{0}/H)^{1/2}$
because the paraconductivity dependence can be valid in the mixed state
without phase coherence only. It is obvious that the phase coherence length
increases in the lower region of the resistive transition where the
current-voltage characteristics become non-Ohmic. We can not determine more
exactly the onset of this transition from the comparison of the
experimental dependencies with the paraconductivity dependence because the
Hartree approximation do not give accurate enough result.

The LLL approximation is valid for the $Nb_{1-x}O_{x}$ films at the
magnetic field values used in our work: $H = 0.1 T \gg
H_{LLL}(Nb_{1-x}O_{x}) \simeq 10^{-4} \ T$. Therefore the deviation of the
excess conductivity of the $Nb_{1-x}O_{x}$ films with strong disorder from
the scaling dependence (Fig.5) is evidence of the increase of the phase
coherence length. This deviation is smooth because the film with strong
pinning disorder is close to the Mendellsohn's model. In our model, in
which thin film with strong disorder is considered as the Mendelssohn
sponge with variable width of superconducting threads $w(T,H) \simeq
a(\Phi_{0}/B)^{0.5} - \xi(T)$, the length of phase coherence increases
primarily as a consequence of the $w(T,H)$ increase. Therefore it depends
not only on the temperature but also of the H value, begins to increase at
$H \simeq H_{c2}$ and increases with T more quickly than in one-dimensional
superconductors with a constant w.

It is interesting that the difference of the excess conductivity dependence
of the film with small grain structure from the scaling law is visible
already above $H_{c2}$ (Fig.5). This means that the length of phase
coherence can exceed $(\Phi_{0}/H)^{1/2}$ already above $H_{c2}$ if pinning
disorder is strong enough. The Mendellsonh model can be valid near
$H_{c2}$ in superconductors with strong pinning disorders. The mean field
critical field of a one-dimensional superconductor (with width w), as well
as the one of a thin film in parallel magnetic field \cite{tinkham},
$H_{c,sp} = 3^{0.5}\Phi_{0}/\pi\xi w$ can exceed $H_{c2} =
\Phi_{0}/2\pi\xi^{2}$. Therefore the phase coherence length of a thin film
with strong disorder can exceed $(\Phi_{0}/H)^{1/2}$ already above
$H_{c2}$.

\

V. CONCLUSIONS

\

The influence of fluctuations on the resistive properties of the mixed
state of layered HTSC $Bi_{2}Sr_{2}CaCu_{2}O_{8+x}$ does not differ
qualitatively from the one of thin films of conventional superconductors.
The appearance of phase coherence is smooth in both cases. Moreover, the
quantitative difference between the behavior of thin films with weak and
strong disorder is greater than the difference between
$Bi_{2}Sr_{2}CaCu_{2}O_{8+x}$ and conventional superconductors. Therefore
the amount of disorder must taken into account in the analysis of
experiments concerning fluctuation effects in thin films and layered
superconductors.

The main difference between the behavior of layered HTSC
$Bi_{2}Sr_{2}CaCu_{2}O_{8+x}$ and of thin films of conventional
superconductors is connected with the used values of magnetic field. The
LLL approximation is valid in this magnetic field region for conventional
superconductors but not for layered HTSC. Strictly speaking, the second
critical field, $H_{c2}$, has a sense only in the LLL approximation,
because the $H_{c2}$ is the field value at which the linearized energy of
the lowest Landau level (but not higher Landau levels) is equal zero
\cite{huebener}. Therefore, one may say that the magnetic field with the
values used in our work, as well as in majority of other works, decreases
the $T_{c2}$ value of conventional superconductors and influences weakly on
the $T_{c}$ value of layered HTSC. The observed difference of the resistive
transition is conditioned by this circumstance: the magnetic field is
enough high in order to displace the onset of the resistive transition of
films of conventional superconductors but not of layered HTSC. It ought be
supposed that this difference will disappear when using higher magnetic
field for $Bi_{2}Sr_{2}CaCu_{2}O_{8+x}$ investigation.

This supposition is confirmed partly by the observed likeness of the
qualitative change of the resistive characteristics. The transition from
the Ohmic current-voltage characteristics to the one with a finite critical
current takes place in the wide intermediate regime both in the
$Bi_{2}Sr_{2}CaCu_{2}O_{8+x}$ and in thin films of conventional
superconductors. This behaviour differs qualitatively from the one observed
in some bulk superconductors with weak pinning disorders where this
transition occurs in a narrow region \cite{nik85}. This means that the
jumping volume of the vortex creep, $V_{j}$, changes by jump in bulk
superconductors with weak pinning and changes weakly with a change of
temperature or magnetic field value in two dimensional superconductors.

The observed changes of the resistive properties in all type II
superconductors placed in a magnetic field can be connected with the change
of the phase coherence length. Four region may be distinguished: 1) the
mixed state without the phase coherence; 2) the transition region
(the $H_{c4}$ region) where the phase coherence length increase to a sample
size; 3) the Abrikosov state with visible thermally activated linear
resistivity; 4) the Abrikosov state with a finite static critical current.
Width and situation of these regions are different in different
superconductors.  First of all the superconductor dimensionality and the
amount of pinning disorders determine the character of the phase coherence
length change.

Experimental investigations show that in bulk superconductors with weak
disorder \cite{nik81l,welp} the sharp deviation of the excess conductivity
value from the scaling law and the appearance of the non-Ohmic
current-voltage characteristic are observed simultaneously, at the same
magnetic field value $H_{c4}$ \cite{nik83,nik85}. This means that the
length of phase coherence changes by jump from $(\Phi_{0}/B)^{1/2}$ to a
sample size L as well as in an ideal superconductor. The $H_{c4}$ region
tightens to a critical point in bulk superconductor with weak disorders.
The regime of the Abrikosov state with visible vortex creep is observed in
some bulk samples and is absent in others \cite{nik85}. But the $H_{c4}$
region is narrow only in few bulk samples. It broadens with the pinning
increase \cite{nik85,fendrich}. Therefore the Mendelssohn's model is more
suitable for the description the long-range phase coherence appearance in
the majority of cases.

The Mendelssohn's model described enough well the changes of the resistive
properties observed in thin films of conventional superconductors with
strong disorders. All four region can be marked out in this case. For
example, in our $Nb_{1-x}O_{x}$ film with small grain structure at T = 4.2
K: 1) the mixed state without the phase coherence exists at $H > 1.05
H_{c2}$, 2) the $H_{c4}$ transition takes place at $0.75 H_{c2} < H < 1.05
H_{c2}$, 3) the thermally activated linear resistivity is observed at $0.2
H_{c2} < H < 0.75 H_{c2}$ and 4) at $H < 0.2 H_{c2}$ the static critical
current is finite. In the wide $H_{c4}$ region the excess conductivity
dependencies deviate from the scaling law but the current-voltage
characteristics remain Ohmic. Below the $H_{c4}$ region the resistive
properties at low current can be described by the Kim-Anderson vortex creep
theory and by the theory of the vortex creep in the Mendelssohn's sponge.
The Abrikosov state with a finite static critical current ought be
considered as the region where the $E_{0}$ value is too small (in the
consequence of the $w(T,H) \simeq 2\xi(T) ((H_{c2}/H) - 1)$ increase) in
order to a resistance can be measured at a low current.

The experimental data obtained in our work do not allow to mark out
completely all four regions in thin films with weak disorders. The scaling
of the excess conductivity dependencies shows that the mixed state without
phase coherence exists at least down to $0.7 T_{c2}$. The current-voltage
characteristics become non-Ohmic in a much lower temperature (magnetic
field value), for example at $0.008 H_{c2}$ and t = 0.75. The $H_{c4}$
transition can take place in the wide region below $0.7 T_{c2}$. We suppose
that the mixed state without phase coherence exists down to very low
magnetic fields.

A sharp feature of the resistive properties should be observed at
the transition to the Abrikosov state in a superconductor with weak pinning
disorders. The feature of the vortex flow resistivity below the transition
is observed both in bulk superconductors \cite{nik81j} and in thin films
\cite{kes97}. The amount of pinning disorders in our amorphous
$Nb_{1-x}O_{x}$ is smaller than in the a-NbGe films used in \cite{kes97}.
Therefore the feature of the resistive dependence should be more sharp in
our films. But no feature is observed. Therefore it is supposed in
\cite{nik95l} that the situation of the transition to the Abrikosov state
is not universal in two-dimensional superconductors, but that it depends on
the amount of pinning disorders and that the Abrikosov state is does not
exist down to very low fields in thin films with very weak disorders. This
supposition is confirmed by a theoretical result obtained in work
\cite{nik95b}.

We can not detect the onset of the transition to the Abrikosov state in
$Bi_{2}Sr_{2}CaCu_{2}O_{8+x}$ from the comparison of the excess
conductivity dependencies with the LLL scaling law, because the LLL
approximation is not valid at $H < Gi H_{c2}(0)$. Results of work
\cite{livan97} allow to suppose that the mixed state without phase
coherence exists in a wide region below $H_{c2}$. The Abrikosov state
exists only at low reduced magnetic fields. Comparison of our results and
results of similar investigations in other works shows that the situation
of the transition to the Abrikosov state depend on the amount of pinning
disorder in $Bi_{2}Sr_{2}CaCu_{2}O_{8+x}$ as well as in thin films of
conventional superconductors. The appearance of pinning in
$Bi_{2}Sr_{2}CaCu_{2}O_{8+x}$ single crystals with weak disorder is
observed in lower magnetic fields than it is observed in our
$Bi_{2}Sr_{2}CaCu_{2}O_{8+x}$ film.

A magnetization step was observed in $Bi_{2}Sr_{2}CaCu_{2}O_{8+x}$ single
crystal in some papers \cite{zeldov95}. This step is interpreted as a
consequence of the vortex lattice melting. This interpretation can not be
right. But this transition can not be interpreted also as the long-range
phase coherence appearance in two-dimensional superconductors because no
sharp transition is observed in thin films \cite{nik95l}. This step can be
connected with a transition from two- to three-dimensional behavior in the
layered superconductor. This transition causes the appearance of the
long-range phase coherence.

The strong increase of the critical current value in low magnetic fields
observed in the amorphous $Nb_{1-x}O_{x}$ is typical for a superconductor
with weak pinning and without weak links. The large value of the critical
current in zero magnetic field is evidence of the absence of weak links.
The small $j_{c}$ value in zero magnetic field and its weak change in low
magnetic fields observed in the $Bi_{2}Sr_{2}CaCu_{2}O_{8+x}$ film can be
connected with presence of weak links.

\

ACKNOWLEDGMENTS

This work was made in the framework of the Project INTAS-96-0452.
We thank the International Association for the Promotion of Co-operation
with Scientists from the New Independent States for financial support.
A.V.N. thanks also the Russian National Scientific Council on
"Superconductivity" of SSTP "ADPCM" (Project 98013) for financial support.

\newpage

Figure Captions

Fig.1. Resistive transitions in perpendicular magnetic field of a) the
amorphous $Nb_{1-x}O_{x}$ film (d = 20 nm) at 1 - H = 0, 2 - H = 0.2 T, 3 -
H = 0.4 T, 4 - H = 0.8 T, 5 - H = 1.2 T; b) the
$Bi_{2}Sr_{2}CaCu_{2}O_{8+x}$ in-plane (along ab) at 1 - H = 0, 2 - H =
0.044 T, 3 - H = 0.2 T, 4 - H = 0.5 T, 5 - H = 1 T; and c) the
$Nb_{1-x}O_{x}$ film with small grain structure (d = 20 nm) at 1 - H = 0, 2
- H = 0.4 T, 3 - H = 1.2 T.

Fig.2. Current-voltage characteristics of a) the amorphous $Nb_{1-x}O_{x}$
film (d = 20 nm) at T = 1.75 K ($T/T_{c} = 0.77$, $H_{c2} = 1.14 T$) and 1
- H = 0, 2 - H = 0.001 T, 3 - H = 0.002 T, 4 - H = 0.004 T, 5 - H = 0.006
T, 6 - H = 0.01 T; b) the $Bi_{2}Sr_{2}CaCu_{2}O_{8+x}$ in-plane (along ab)
at T = 60 K ($T/T_{c} = 0.75$, $H_{c2} = 20 T$) and 1 - H = 0, 2 - H = 0.02
T, 3 - H = 0.05 T, 4 - H = 0.1 T, 5 - H = 0.2 T, 6 - H = 0.4 T; c) the
$Nb_{1-x}O_{x}$ film with small grain structure (with d = 20 nm) at T = 4.2
K ($T/T_{c} = 0.74$, $H_{c2} = 0.9 T$) and 1 - H = 005, 2 - H = 0.1 T, 3 -
H = 0.2 T, 4 - H = 0.4 T, 5 - H = 0.5 T, 6 - H = 0.7 T.

Fig.3. The static critical current $j_{cs}$ dependencies on the reduced
magnetic field $H/H_{c2}$ of the amorphous $Nb_{1-x}O_{x}$ film at T = 1.75
K (t = 0.76) (curve 1); the $Bi_{2}Sr_{2}CaCu_{2}O_{8+x}$ at T = 60 K (t =
0.75) (curve 2); the $Nb_{1-x}O_{x}$ film with small grain structure at T =
4.2 K (t = 0.74) (curve 3). The $j_{cs}$ values were determined by the
voltage level 0.0001 V/m.  $H/H_{c2} = 10^{-4}$ is corresponded to $H
\simeq 0$.  $j_{cs} = 10^{4} \ A/m^{2}$ is correspond to $j_{c} \simeq 0$.

Fig.4. Current-voltage characteristics of the $Nb_{1-x}O_{x}$ film with
small grain structure (with d = 20 nm) at T = 4.2 K and H = 0.6 T (curve 1)
and H = 0.4 T (curve 2), the amorphous $Nb_{1-x}O_{x}$ film at T = 1.75 K
and H = 0.004 T (curve 3) and the $Bi_{2}Sr_{2}CaCu_{2}O_{8+x}$ at T = 60 K
and H = 0.05 T (curve 4). The lines $1^{*}$, $2^{*}$ and $3^{*}$ denote the
$E = E_{0}sinh(j/j_{0})$ dependencies with $1^{*}$ - $E_{0} = 0.27 \ V/m$
and $j_{0} = 0.90 \ 10^{8} \ A/m^{2}$; $2^{*}$ $E_{0} = 0.0015 \ V/m$ and
$j_{0} = 0.43 \ 10^{8} \ A/m^{2}$; $3^{*}$ $E_{0} = 0.050 \ V/m$ and $j_{0}
= 0.86 \ 10^{8} \ A/m^{2}$.

Fig.5. The $\Delta \sigma (Gi_{2D}ht)^{0.5}d/t\sigma_{0}$ versus
$(t-t_{c2})/(Gi_{2D}ht)^{0.5}$ dependencies of the amorphous
$Nb_{1-x}O_{x}$ film ($T_{c} = 2.37 \ K$; $Gi_{2D} \simeq 0.0005$)
at H = 0.1 T (curve 1), H = 0.4 T (curve 2), H = 1.2 T (curve 3);
the $Nb_{1-x}O_{x}$ film with small grain structure ($T_{c} = 5.7 \ K$;
$Gi_{2D} \simeq 0.0001$) at H = 0.4 T (h = 0.11) (curve 4), H
= 1.2 T (h = 0.33) (curve 5) and the $Bi_{2}Sr_{2}CaCu_{2}O_{8+x}$ ($T_{c}
= 79 \ K$; $Gi_{2D} \simeq 0.02$) at H = 0.1 T (h = 0.0012) (curve 6), H =
1 T (h = 0.012) (curve 7).


\begin{thebibliography}{9}

\bibitem{blatter} G.Blatter, M.V.Feigel'man, V.B.Geshkenbein, A.I.Larkin,
and V.M.Vinokur, Rev.Mod.Phys. {\bf66}, 1125 (1994).

\bibitem{fisher} D.S.Fisher, M.P.A.Fisher, and D.A.Huse, Phys.Rev.
B {\bf 43}, 130 (1991).

\bibitem{huebener} R.P.Huebener, Magnetic Flux Structures in
         Superconductors (Springer-Verlag, Berlin Heidelberg New York, 1979).

\bibitem{salamon} M.B.Salamon, "Thermodynamic Properties, Fluctuations
and Anisotropy of High-Tc Superconductors", in Physical Properties of High
Temperature Superconductors I, Ed.  D.M.Ginsberg (World Scietific,
Singapure, 1989) p.39.

\bibitem{lee72} P.A.Lee and S.R.Shenoy, Phys.Rev.Lett. {\bf 28}, 1025
         (1972).

\bibitem{nik96} A.V.Nikulov, "Fluctuation Effects in Mixed State of Type
II Superconductors", in Fluctuation Phenomena in High Temperature
Superconductors (Ed.  M.Aussloos and A.A.Varlamov) (Kluwer Academic
Publishers, Dordrecht/Boston/London, 1997) p.271.

\bibitem{abrikos} A.A.Abrikosov, Zh.Eksp.Teor.Fiz. {\bf32}, 1442 (1957)
(Sov.Phys.-JETP {\bf5}, 1174  (1957) ).

\bibitem{nik98} A.V.Nikulov, submitted to Phys.Rev.;
http://xxx.lanl.gov/abs/cond-mat/9812168

\bibitem{hake} Farrant, S.P. and Gough C.E., Phys.Rev.Lett. {\bf34},
943 (1975); R.F.Hassing, R.R.Hake, and L.J.Barnes, Phys.  Rev.Lett.
{\bf30}, 6 (1973).

\bibitem{nik84} V.A.Marchenko and A.V.Nikulov, Zh.Eksp.Teor.Fiz. {\bf86},
1395 (1984) (Sov.Phys.-JETP {\bf59}, 815 (1984)).

\bibitem{chen} D.-X.Chen, J.J.Moreno, A.Hernando, A.Sauchez, and B.-Z.Li,
Phys.Rev.B {\bf 57}, 5059 (1998).

\bibitem{tinkham} M.Tinkham, Introduction to Superconductivity
(McGraw-Hill Book Company, 1975)

\bibitem{nik97} A.V.Nikulov, S.V.Dubonos, and Y.I.Koval, J.Low Temp.Phys.
{\bf 109}, 643 (1997).

\bibitem{nik81l} V.A.Marchenko and A.V.Nikulov, Pisma
Zh.Eksp.Teor.Fiz. {\bf34}, 19 (1981) (JETP Lett. {\bf34}, 17 (1981)).

\bibitem{welp} W.K.Kwok, U.Welp, G.W.Crabtree, K.G.Vandervoort,
R.Hulscher, and J.Z.Liu, Phys.Rev.Lett. {\bf64}, 966 (1990); H.Safar,
P.L.Gammel, D.A.Huse, D.J.Bishop, J.P.Rice, and D.M.Ginzberg,
Phys.Rev.Lett. {\bf69}, 824 (1992); W.K.Kwok, S.Fleshler, U.Welp,
V.M.Vinokur, J.Downey, G.W.Crabtree, and M.M.Miller, Phys.Rev.Lett.
{\bf69}, 3370 (1992); W.Jiang, N.-C.Yeh, D.S.Reed, U.Kriplani, and
F.Holtzberg, Phys.Rev.Lett. {\bf74}, 1438 (1995).

\bibitem{nik90} A.V.Nikulov, Supercond.Sci.Technol. {\bf3}, 377 (1990).

\bibitem{nik99} A.V.Nikulov, "Vortex Lattice Melting Theories as an Example
of Science Fiction", in Symmetry and Pairing in Superconductors (Eds.
M.Aussloos and S.Kruchinin) (Kluwer Academic Publishers, 1999) p.131;
http://xxx.lanl.gov/abs/cond-mat/9811051

\bibitem{shilling} A.Schilling et al., Nature {\bf 382}, 791 (1996).

\bibitem{mendelss} K.Mendelssohn, Proc.Roy.Soc. {\bf152A}, 34 (1935).

\bibitem{grunberg} L.W.Grunberg and L.Gunther, Phys.Lett. A {\bf38},
463 (1972).

\bibitem{kim} Y.B.Kim, C.F.Hempsted, A.R.Strnad, Phys.Rev. {\bf131},
2486 (1963); Phys.Rev. {\bf139}, A1163 (1965).

\bibitem{anders64} P.W.Anderson and Y.B.Kim, Rev.Mod.Phys. {\bf 36}, 39
(1964).

\bibitem{brandt95} E.H.Brandt, Rep.Progr.Phys. {\bf58}, 1465 (1995).

\bibitem{ami78} S.Ami and K.Maki, Phys.Rev.B {\bf18}, 4714 (1978).

\bibitem{1964} Kleiner, W.H., Roth, L.M., and Autler S.H., Phys.Rev. A
{\bf133}, 1226 (1964).

\bibitem{macdonal} J.Hu and A.H.MacDonald, Phys.Rev. B {\bf 52}, 1286
(1995)

\bibitem{nik95b} A.V.Nikulov, Phys.Rev. B {\bf52}, 10429 (1995) .

\bibitem{thouless} D.J.Thouless, Phys.Rev.Lett. {\bf34}, 946 (1975).

\bibitem{nik83} V.A.Marchenko and A.V.Nikulov, Fiz.Nizk.Temp. {\bf9},
816 (1983).

\bibitem{nik85} A.V.Nikulov, Thesis, Institute of Solid State Physics,
Chernogolovka, 1985.

\bibitem{nik95l} A.V.Nikulov, D.Yu.Remisov, and V.A.Oboznov,
Phys.Rev.Lett. {\bf75}, 2586 (1995).

\bibitem{kes97} M.H.Theunissen and P.H.Kes, Phys.Rev.B {\bf55}, 15183
(1997).

\bibitem{bmmpp91} G.Balestrino, M. Marinelli, E. Milani, A. Paoletti, and
P. Paroli, J.Appl.Phys. {\bf70}, 6939 (1991).

\bibitem{kes93} P.Berghuis and P.H.Kes, Phys.Rev.B {\bf47}, 262
(1993).

\bibitem{kopnin75} L.P.Gor'kov and N.B.Kopnin, Usp.Fiz.Nauk {\bf116},
413 (1975) (Sov.Phys. - Uspechi {\bf18}, 496 (1976)).

\bibitem{livan97} D.V.Livanov, E.Milani, G.Balestrino, and C.Aruta,
Phys.Rev.B {\bf55}, 8701 (1997).

\bibitem{nik81j} V.A.Marchenko and A.V.Nikulov, Zh.Eksp.Teor.Fiz.
{\bf80}, 745 (1981) (Sov.Phys.-JETP {\bf53}, 377 (1981)).

\bibitem{fendrich} J.A.Fendrich et al., Phys.Rev.Lett. {\bf74}, 1210
(1995).

\bibitem{zeldov95} E.Zeldov et al., Nature {\bf375}, 373 (1995)

\end{thebibliography}
\end{document}